\documentclass[prd,twocolumn,showpacs,preprintnumbers,amsmath,amssymb,nofootinbib,APS,10pt]{revtex4-1}

\usepackage{dcolumn}
\usepackage{bm}
\usepackage[applemac]{inputenc}
\usepackage[spanish,english]{babel}
\usepackage{amsmath,amssymb,amsfonts,latexsym,cancel}
\usepackage{graphicx}
\usepackage{color}
\usepackage{hyperref}

\newcommand*{\Scale}[2][4]{\scalebox{#1}{$#2$}}%
\newcommand{\be}{\begin{equation}}
\newcommand{\ee}{\end{equation}}
\newcommand{\bea}{\begin{eqnarray}}
\newcommand{\eea}{\end{eqnarray}}

\begin{document}
\title{{\bf Spacetime correlators of perturbations in slow-roll de Sitter inflation}}

\author{Adrian del Rio and  Jose Navarro-Salas}
\affiliation{ {\footnotesize Departamento de Fisica Teorica and
IFIC, Centro Mixto Universidad de Valencia-CSIC.
    Facultad de Fisica, Universidad de Valencia,
        Burjassot-46100, Valencia, Spain }}

\date{12, March 2014}

\begin{abstract}

Two-point correlators and self-correlators   of primordial perturbations in  quasi-de Sitter spacetime backgrounds are considered. For large separations two-point correlators exhibit nearly scale invariance, while for short distances self-correlators need standard renormalization. We study the deformation of two-point correlators to smoothly match the self-correlators at coincidence. The corresponding angular power spectrum is evaluated in the Sachs-Wolfe regime of low multipoles. Scale invariance is maintained,   but the amplitude of $C_{\ell}$  could change in a nontrivial way.\\

\end{abstract}


\date{March 13, 2014}

\maketitle

\section{Introduction} 
In recent years de Sitter space has received considerable attention. Astronomical observations \cite{Riess-Perlmutter} are pointing out that our universe now has a very tiny positive cosmological constant, which, however, embodies around three quarters of the energy of the observable universe. Moreover, according to inflationary cosmology the very early universe underwent a period of very rapid expansion powered by a large effective cosmological constant. The discovery of the anisotropies in the cosmic microwave background (CMB) \cite{cobe} constitutes a very sensitive probe of the primordial density perturbations and its quantum mechanical origin \cite{inflation2}. The comparison of  observations \cite{planck13} with theoretical predictions is currently a sharp tool to test inflation and the theory of quantized fields in curved backgrounds \cite {parker-toms, birrell-davies}. Therefore, a precise understanding of the quantum properties of fields in de Sitter space is fundamental
  for
  both the very early and the late-time universe. 

In this article we  will focus on the quantum treatment of primordial perturbations, which will be regarded as quantum fields $\phi$ living in a curved (quasi-de Sitter) spacetime. The two-point correlation function $\langle \phi (t, \vec x) \phi (t, \vec x')\rangle$ exhibits scale invariance at large separations $|\vec x - \vec x'| H \gg 1$ or, equivalently, at late-time $Ht \gg 1$. On the other hand, the amplitude of the perturbation at a given spacetime point  can be quantified by the self-correlator $\langle \phi^2(t, \vec x)\rangle$, which requires one to get rid off the corresponding ultraviolet (UV) divergences and renormalize the expectation value.

In the first part of this work we will compute and analyze the above quantities in a slow-roll de Sitter background. We will also project the large-distance behavior of the correlator of scalar perturbations on a sphere of fixed radius. This sphere is linked by time evolution to the last scattering surface, where the cosmic microwave background and its anisotropies are formed, and the  angular power spectrum can easily be  obtained within this spacetime picture.  In the second part we will 
redo the calculation for the angular power spectrum by using a deformed two-point correlator. The new correlator is defined in such a way that it matches the self-correlator at coincidence. To this end we naturally use methods of renormalization in homogeneous backgrounds \cite{parker-toms, birrell-davies}. The revised angular power spectrum maintains the nearly scale invariance, but the amplitude of the multipole coefficients $C_{\ell}$ may be altered in a non-trivial way. We will focus on the lower multipoles, where the Sachs-Wolfe effect dominates.

\section{Spacetime correlators  in slow-roll inflation} 

\subsection{Correlator of tensorial perturbations in a slow-roll scenario}

Tensorial perturbations can be described by  two independent, massless scalar fields  propagating in the unperturbed quasi de Sitter background. These two scalar fields represent the two independent polarization components of the fluctuation tensorial modes $\mathcal D_{ij}$ in the  inflationary universe: $ ds^2= dt^2 -a^2(t)(\delta_{ij} + \mathcal D_{ij})dx^idx^j$. Expanding the fluctuating fields $\mathcal D_{ij}$ in plane wave modes $\mathcal D_k(t) e_{ij}e^{i\vec{k}\vec{x}}$,
where $e_{ij}$ is a constant polarization tensor obeying the conditions $e_{ij}=e_{ji}, e_{ii}=0$ and $k_ie_{ij}=0$,
one obtains the equation
$
\ddot{\mathcal D}_k + 3H \dot{\mathcal D}_k +\frac{k^2}{a^2} \mathcal D_k=0 $,
with $k\equiv |\vec{k}|$ and $H=\dot a/a$.
The conditions for the polarization tensor imply that the
perturbation field $\mathcal D_{ij}$ can be decomposed into two polarization
states described by a couple of  massless scalar fields $\mathcal D_{ij}=\mathcal D_+ e_{ij}^++\mathcal D_{\times}e_{ij}^{\times}$, where $e_{ij}^{s}e_{ij}^{s'}=2\delta_{ss'}$ ($s=+,\times$ stands for the two independent polarizations), both obeying the above wave equation 
 (see, for instance, \cite{Weinberg}). For simplicity  we omit the subindex $+$ or $\times$.

In the slow-roll approximation one assumes that the Hubble parameter $ H(t)$ changes very gradually, and the change is parametrized by a slow-roll parameter $\epsilon \equiv -\dot{H}/H^2 \ll 1$.  Within this approximation it is possible to solve the wave equation in a closed form in terms of the conformal time 
$\eta \equiv\int dt/a(t)$. Taking into account that $(1-\epsilon)\eta= -\frac{1}{aH}$, the wave equation for $\mathcal D_k$ turns out to be of the form 
\be 
\frac{d^2 \mathcal D_k}{d\eta^2} -\frac{2}{\eta(1-\epsilon)}\frac{d \mathcal D_k}{d\eta}+ k^2\mathcal D_k=0 \ .\label{6}
 \ee
Treating now the parameter $\epsilon$ as a constant, one can  univocally solve the above equation with the requirement of recovering, for $\epsilon \to 0$,  the Bunch-Davies vacuum \cite{BD}. The properly normalized solutions for the modes are
\be
 \mathcal D_{{k}}(t) = \frac{\sqrt{16\pi G}}{\sqrt{2(2\pi)^3 a^3}}(-\eta\, a\,
\pi /2)^{1/2}H^{(1)}_{\nu}(-k\eta)   \label{7} \ , 
\ee
where $G$ is the Newton constant and  the index of the Hankel function is  exactly 
$\nu=\frac{3}{2} + \frac{\epsilon}{1-\epsilon}$.
Having the explicit form of the modes, we can now compute the two-point function. At equal times $t=t'$ we find
\bea 
\label{8}\langle \mathcal D (t, \vec{x}) \mathcal D(t, \vec{x}') \rangle&=& \frac{G}{\pi a^2\eta^2} \Gamma\left(\frac{3}{2}+\nu\right) \Gamma \left(\frac{3}{2}-\nu\right) \nonumber \\ &\times&{_{2}}F_1\left(\frac{3}{2}+\nu, \frac{3}{2}-\nu; 2; 1-\frac{(\Delta x)^2}{4\eta^2}\right) \ . 
\eea
where $ \Delta x \equiv |\vec x - \vec x'|$. For $\nu=3/2$ ($\epsilon =0$) we  have the unavoidable  infrared  divergence of the Bunch-Davies vacuum \cite{IR}.

For large separations  $ a \Delta x \gg H^{-1}$ one obtains
\be
 \label{9}\langle \mathcal D (t, \vec{x}) \mathcal D (t, \vec{x}') \rangle \sim \frac{4G \Gamma(3/2-\nu)}{\pi^{3/2}}\frac{\Gamma(\nu)}{a^2\eta^2} \left (\frac{\Delta x}{-\eta}\right )^{2(\nu - 3/2)} \ . 
 \ee
One can immediately observe that the amplitude above is nearly scale invariant. Moreover, the term $(-\eta)^{1-2\nu}/a^2$ is time independent, which allows us to evaluate it at the most convenient time. In fact, the correlator can be rewritten as
\be 
\label{10}  \langle \mathcal D (t, \vec{x}) \mathcal D (t, \vec{x}') \rangle \sim  -\frac{16 \pi G}{2\epsilon} \left (\frac{H(t_{\Delta x})}{2\pi }\right )^{2}  
\ , \ee 
where the time $t_{\Delta x}$ is defined as $a(t_{\Delta x}) \Delta x = H^{-1}(t_{\Delta x})$. 
Note  that there is an implicit $\Delta x$ dependence on $H(t_{\Delta x})$, given by the one in (\ref{9}).

\subsection{Correlator of scalar perturbations in a slow-roll scenario} 
Scalar perturbations can be studied through the gauge-invariant field
$\cal{R}$ (the comoving curvature perturbation; see, for instance, \cite{Weinberg}).
For single-field inflation, the modes of the scalar perturbation are given by
  \be 
  \label{12} {\cal{R}}_k( t) = (-\pi \eta /4(2\pi)^3 z^2)^{1/2}H^{(1)}_{\nu}(-\eta k) \ ,
   \ee   
where now $\nu= 3/2 + (2\epsilon +\delta)/(1-\epsilon)$ and $\delta \equiv \ddot H/2H\dot H$ is a second slow-roll parameter. Moreover, $z\equiv a\dot{\phi}_0/H$, where $\phi_0(t)$ is the homogeneous part of the inflaton field.
These modes determine the vacuum state of scalar perturbations. Such a state can also be regarded as the natural extension of the Bunch-Davies vacuum of de Sitter space. The corresponding two-point function $\langle {\cal R} (t, \vec {x}), {\cal R} (t, \vec {x}') \rangle$ is given by
\bea
 \label{2pfR}\langle {\cal R} (t, \vec {x}) {\cal R} (t, \vec {x}') \rangle  &=& \frac{1}{16 \pi^2 z^2\eta^2} \Gamma\left(\frac{3}{2}+\nu\right) \Gamma \left(\frac{3}{2}-\nu\right)  \\ &\times&{_{2}}F_1\left(\frac{3}{2}+\nu, \frac{3}{2}-\nu; 2; 1-\frac{(\Delta x)^2}{4\eta^2}\right)  \ . \nonumber 
\eea

For separations larger than the Hubble radius $a|\vec{x}-\vec{x}'| \gg H^{-1}$
we get   
\be
 \label{2pfR2}\langle {\cal R} (t, \vec {x}) {\cal R} (t, \vec {x}') \rangle \sim \frac{\Gamma(\frac{3}{2}-\nu) }{4 \pi^2 z^2\eta^2} \frac{\Gamma (\nu)}{\sqrt{\pi}} \left (\frac{\Delta x}{-\eta}\right )^{2(\nu - 3/2)} \ . 
\ee
This expression can be rewritten, assuming $\nu - 3/2\equiv (1-n)/2 \approx 0$ ($n$ is  the scalar spectral index), as  
\be
 \label{13}\langle {\cal R} (t, \vec {x}) {\cal R} (t, \vec {x}') \rangle \sim  -\frac{4\pi G}{(1-n)  \epsilon} \left (\frac{H(t_{\Delta x})}{2\pi }\right )^{2} \ . \ee   
 
 \subsection{Angular power spectrum}

Restricting the two-point function  of scalar perturbations to points such that $|\vec{x}|=|\vec{x}'|$, we can further obtain $
\Delta x^{1-n} = 2^{\frac{1-n}{2}}|\vec{x}|^{(1-n)}(1-\cos \theta)^{(1-n)/2}$ where $\theta$ is the angle formed by $\vec{n}=\vec{x}/|\vec{x}|$ and $\vec{n}'=\vec{x}'/|\vec{x}|$. Then, taking $|\vec{x}|= r_L$, where $r_L$ is the comoving radial coordinate of the last scattering surface
\bea
r_L= H(t_0)^{-1}a(t_0)^{-1}\int_{\frac{1}{1+z_L}}^1 \frac{dx}{\sqrt{\Omega_{\Lambda} x^4+\Omega_M x+\Omega_R}} \ , \ \ \ 
\eea
with the standard cosmological values for $z_L$, $\Omega_{\Lambda}$, $\Omega_{R}$, and $\Omega_M$ \cite{planck13},
 the correlator of scalar perturbations for large separations (\ref{2pfR2}) shows exactly
\bea
 \label{15a}\langle {\cal R} (t, \vec {x}) {\cal R} (t, \vec {x}') \rangle & \sim & \frac{4\pi G}{\epsilon} \frac{H^2(1-\epsilon)^2}{16\pi^2}\frac{4\Gamma\left(2-\frac{n}{2}\right)}{\sqrt{\pi}} \Gamma\left(\frac{n-1}{2}\right)   \nonumber\\
 & \times & 2^{\frac{1-n}{2}}\bar r_L^{1-n} \left (1-\cos{\theta} \right )^{\frac{1-n}{2}} 
\eea
where we have defined the dimensionless quantity $\bar{r}_L(t)\equiv H(1-\epsilon)\, a r_L$. 
This two-point function can be  related to the temperature fluctuations in the CMB, 
\be
 \langle \Delta T (\vec{n}) \Delta T (\vec{n}') \rangle = \sum_{\ell} C_{\ell} \frac{2\ell+1}{4\pi} P_{\ell}(\cos \theta) \ , \label{16}
 \ee
where $P_{\ell}$ is the Legendre polynomial, via the Sachs-Wolfe effect (see, e.g., \cite{Weinberg}) $\langle \Delta T (\vec{n}) \Delta T (\vec{n}') \rangle_{SW}= \frac{T_0^2}{25} \langle {\cal R} (r_L\vec {n}) {\cal R} (r_L\vec {n}') \rangle $. The coefficients $C_{\ell}$ are obtained by Legendre transformation of (\ref{16}),
\bea
\Scale[0.97]{
C_{\ell}^{SW}=\frac{2\pi T_0^2}{25}\int_{-1}^{1}d\cos{\theta} P_{\ell}(\cos{\theta}) \langle {\cal R} (r_L\vec {n}) {\cal R} (r_L\vec {n}') \rangle } \ .  \label{16a}
\eea
Therefore, the low multipole coefficients, dominated by the Sachs-Wolfe effect, are proportional to the integral
\be 
\label{17}C^{SW}_{\ell} \propto \int_{-1}^{1} dy (1-y)^{\frac{1-n}{2}} P_{\ell}(y) \ , 
\ee
with $y\equiv \cos\theta$. The above integral can be computed analytically \cite{Tablas, handbook}, and we finally find  
\bea
C^{SW}_{\ell}&=&\frac{8\pi T_0^2}{25}\frac{4\pi G}{\epsilon} \frac{H^2(1-\epsilon)^2}{16\pi^2}\frac{\Gamma(3-n)\Gamma(\ell+\frac{n-1}{2})}{\Gamma(\ell+2-\frac{n-1}{2})} \bar r_L^{1-n}\ , \nonumber \\ \label{19b}
\eea
in exact agreement with the result  obtained with the  momentum-space power spectrum $P_s(k)=|N|^2k^{n-1}$ \cite{Weinberg},
\be
 \label{19c}C_{\ell, SW} = \frac{16\pi^2 T_0^2}{25}\int_0^{\infty} \frac{dk}{k}P_{s}(k) j^2_{\ell}(kr_L) \ , 
 \ee
 where the amplitude $|N|^2$ is given by $|N|^2= \frac{4\pi G}{\epsilon} \frac{H^2(1-\epsilon)^2}{16\pi^2}\frac{ 2^{3-n}}{\pi^2}\Gamma\left(2-\frac{n}{2}\right)^2 \left(\frac{\bar r_L}{r_L}\right)^{1-n}
 \sim \frac{8\pi G H^2}{32\pi^3 \epsilon}$.
For completeness, taking approximately $n\approx 1$ in  (\ref{19b}) and using the standard  assumption $\bar r_L^{(1-n)} \approx O(1) $ \cite{Weinberg}, the estimated order of magnitude for the amplitude of $C_{\ell}^{SW}$ is
\bea  
\ell(\ell+1)C_{\ell}^{SW}  \approx   \frac{2G H^2 T_0^2}{25\epsilon}\bar r_L^{1-n}\sim  \frac{2G H^2 T_0^2}{25\epsilon} \ . \label{estandar}
\eea
 We will go back to this point at the end of Sec. IV.
 
 We note that if the coefficients $C_{\ell}^{SW}$ in (\ref{16a}) were actually evaluated using the exact expression (\ref{2pfR}) for the two-point function 
$ \langle {\cal R} (t, \vec {x}) {\cal R} (t, \vec {x}')\rangle $, the integral (\ref{16a}) would have been divergent [due to the UV divergences of  (\ref{2pfR}) as  points $\vec{x}$ and $\vec x'$ merge]. The use of the large distance behavior   (\ref{2pfR2} and \ref{15a}) everywhere in the integral (\ref{16a}) bypasses the UV divergences and makes the integral convergent. 
 We will see in Sec. IV how the use of a renormalized form of the two-point correlator $ \langle {\cal R} (t, \vec {x}) {\cal R} (t, \vec {x}')\rangle $ does the same job, but with a  slightly different final result for the integral. In a certain limit both results eventually agree, but in general we  find a difference that could be potentially probed   by observations.

\section{Amplitude of quantum perturbations}

We shall use $\phi$ to denote both scalar and tensorial fluctuations, and $\nu$ to represent the corresponding Hankel index.  The  two-point function $\langle \phi(t, \vec x), \phi(t, \vec x')\rangle$ can be expanded at short distances as 
 \begin{widetext}
\bea 
\Scale[1]{
\langle \phi (t, \vec x) \phi (t, \vec x') \rangle= \frac{ H^2(1-\epsilon)^2}{16\pi^2} \left\{  \frac{4}{\Delta \bar{x}^2} +\left(\frac{1}{4}-\nu^2\right)\left(-1+2\gamma+\psi(3/2-\nu)+\psi(3/2+\nu)+\log \frac{\Delta \bar{x}^2}{4}\right) + O(\Delta \bar{x}^2)\right\}   }
 \label{40}
 \ , \eea 
 where we have introduced  the dimensionless quantity $\Delta \bar{x}\equiv H(1-\epsilon)\, a \Delta x$. An additional prefactor, $4\pi G/\epsilon$ or $16\pi G$, needs to be included in considering scalar or tensorial perturbations, respectively. As expected, one encounters the typical quadratic and logarithmic 
 UV divergences of a quantum field in a curved background. Since we are now interested in evaluating the mean square fluctuation
 $\langle \phi^2 (t, \vec x) \rangle$ at a given spacetime point, we have to remove these divergences by standard renormalization in curved spacetime \cite{birrell-davies, parker-toms}.  Different methods can be used to this end.
 A preferred method for our purposes is the point-splitting version of the adiabatic regularization scheme \cite{parker-fulling,  parker-toms,  landete}. In short, the two-point function $\langle \phi (t, \vec x) \phi (t, \vec x') \rangle$ at coincidence can be naturally renormalized by subtracting the second-order adiabatic terms $G_{Ad}^{(2)}((t, \vec x), (t, \vec x'))$
 \bea
 \langle \label{2pfr} \phi (t, \vec x) \phi (t, \vec x') \rangle_{ren}\equiv \langle \phi (t, \vec x) \phi (t, \vec x') \rangle-G_{Ad}^{(2)}((t, \vec x), (t, \vec x')) \label{44}
 \eea 
  and taking the limit $\vec x'\to \vec x$. The method determines univocally the subtraction terms, which are found to be 
 \small
 \bea
G_{Ad}^{(2)}((t, \vec x), (t, \vec x'))= \frac{ H^2(1-\epsilon)^2}{16\pi^2}\left\{   \frac{4}{\Delta\bar{x}^2}+\left(\frac{1}{4}-\nu^2\right) \log{\frac{\Delta\bar{x}^2}{4}} +\frac{2-\epsilon}{3(1-\epsilon)^2}+\left(\frac{1}{4}-\nu^2\right) \left(2\gamma+\log{\frac{\mu^2}{H^2(1-\epsilon)^2}}\right) \right\}  \label{59}
\ , \eea 
\normalsize
where $\mu$ is a renormalization scale and  the corresponding prefactor mentioned above for scalar or tensorial perturbations must be considered \cite{agullo-navarro}.  We observe immediately that the UV divergences cancel  exactly and  we are left  with 
 \be  \langle {\phi}^2 (t, \vec {x}) \rangle_{ren} = \frac{H^2(1-\epsilon)^2}{16\pi^2} \left\{ \left(\frac{1}{4}-\nu^2\right)\left( -1+\psi(3/2-\nu)+\psi(3/2+\nu)-\log \frac{\mu^2}{H^2(1-\epsilon)^2}\right) -\frac{2-\epsilon}{3(1-\epsilon)^2}  \right\} \ . \label{60} \ee  
 \end{widetext}
The above self-correlators quantify the amplitude of perturbations at a given spacetime point.

\section{Modified  correlators  and  angular power spectrum}

In previous sections we have studied the correlator $\langle \phi(t, \vec x) \phi (t, \vec x')\rangle$ and self-correlator $\langle \phi^2(t, \vec x)\rangle$ of tensorial and scalar perturbations in slow-roll inflation.  For an ordinary quantum mechanical system, with a finite number $N$ of degrees of freedom, expectation values of the form  $\langle \phi (i) \phi (j) \rangle$ and $\langle \phi^2(i)\rangle$ match when $j=i$ [for instance, in a chain of spins with $\phi (i) \equiv S_z(i)$ and $i=1, \dots, N$].  However, we are  facing here a field theory (with an infinite number of degrees of freedom), and the above matching  is not {\it a priori} guaranteed. This is so because the self-correlator requires renormalization. We may either assume  this discontinuity  or  modify the two-point correlation function to force  it  to match $\langle \phi^2(t, \vec x)\rangle_{ren}$ in the coincidence limit $\vec x' \to \vec x$ \cite{parker-toms}.  This second possibility was indirectly explored in 
 \cite{parker07} by 
 analyzing the power spectrum of perturbations in  momentum space. 
 It has been somewhat debated in the literature and  properly reviewed in  \cite{Bastero-Gil}.
 One could naturally argue that the point-separated correlator has a well-defined definition in the distributional sense and there is not a mathematical need for any regularization \cite{Bastero-Gil, Finelli}.  However, as the spatial points approach each other, the two-point correlator will grow without bound and diverge as the points merge. Therefore, from the physical point of view it seems reasonable to use a regularized form of the two-point correlator to consistently match the self-correlator at coincidence \cite{parker-toms, note2}. In the conventional approach the expectation value of the self-correlator $\langle \phi^2(t, \vec x) \rangle$ plays almost no role. 
 We assume here that the (renormalized) self-correlator is actually playing a physical role (as in  the Casimir effect). As we will see shortly, the regularized form of the two-point correlator makes the integral (\ref{16a}) UV convergent.  The consequences of this merit  to be explored. 
 Therefore, we  further analyze here this possibility taking advantage of the spacetime viewpoint sketched above. 

We shall modify the correlators by adding the subtraction terms prescribed by renormalization and according to (\ref{2pfr}). We note that a distinguishing characteristic of adiabatic renormalization is that the subtraction terms $G_{Ad}^{(2)}((t, \vec x), (t, \vec x'))$ are well-defined for arbitrary point separation. In general this is  not  possible for an arbitrary spacetime, but for the homogeneous spaces relevant in  cosmology the adiabatic subtraction terms extend to arbitrary large distances.  With this in mind, we will finally compute the angular power spectrum for primordial perturbations using the modified spacetime correlators.

As a previous step we will compute the two-point function at leading order in slow-roll. 
 
\subsection{Two-point function at leading order in slow-roll} 

The procedure is similar for scalar and tensorial fluctuations, so we will do a general treatment. 
First, we start off splitting Eqs. (\ref{8}) and (\ref{2pfR}) as a combination of two hypergeometric functions. To this end we use the transformation properties of hypergeometric functions  \cite{handbook}
\begin{widetext}
\bea
F\left(\frac{3}{2}+\nu,\frac{3}{2}-\nu,2,1-Z \right) & = &  
\frac{Z^{-\frac{3}{2}-\nu}\Gamma{(-2\nu)}} {\Gamma{(\frac{3}{2}-\nu)}\Gamma{(\frac{1}{2}-\nu})}Re\left\{F\left(\frac{3}{2}+\nu,\frac{1}{2}+\nu,1+2\nu,\frac{1}{Z} \right)\right\}\nonumber\\
& + &  \frac{Z^{-\frac{3}{2}+\nu}\Gamma{(2\nu)}} {\Gamma{(\frac{3}{2}+\nu)}\Gamma{(\frac{1}{2}+\nu})}Re\left\{F\left(\frac{3}{2}-\nu,\frac{1}{2}-\nu,1-2\nu,\frac{1}{Z} \right)\right\}    \label{48} 
 \eea
with $Z=\Delta \bar x^2/4\geq 0$.
 We  now  expand  expression (\ref{48}) as a power series of the ``slow-roll'' parameter $\nu$ around $\nu=3/2$, and stay at  first order (for details see the Appendix).  Grouping  terms,  we arrive at the following expression for the two-point function:
\bea
\Scale[1]{
 \langle \phi (x) \phi (x') \rangle \approx \frac{H^2(1-\epsilon)^2}{16\pi^2}\left\{\frac{4}{\Delta \bar{x}^2}-2\log{\frac{\Delta \bar{x}^2}{4}}-1 \right.  +  \frac{2}{(3/2 - \nu)} \Big(\frac{\Delta\bar x^2}{4}  \Big)^{\nu-3/2}   
  +  \left. 4Re\left[ \log\left( \frac{\Delta \bar{x}}{2}+\sqrt{\frac{\Delta \bar{x}^2}{4}-1}\right)\right] \right\}    } \label{approx}
 \eea 
Notice that the UV divergences are just the same as those found in (\ref{40}), but now they are obtained at leading order in the slow-roll expansion. We recover exactly expression (\ref{40}) taking the limit $\Delta \bar x\rightarrow 0$ and the slow-roll approximation.
 
 \subsection{Modified two-point function}
 We can now proceed to do the subtraction. The modified two-point function  then reads 
\bea
\Scale[1]{
 \langle \phi (x) \phi (x') \rangle_{ren} \approx  \frac{H^2(1-\epsilon)^2}{16\pi^2} \left\{\frac{2}{(3/2 - \nu)}\Big(\frac{\Delta\bar x^2}{4}  \Big)^{\nu-3/2} +4Re\left[ \log\left( \frac{\Delta \bar{x}}{2}+\sqrt{\frac{\Delta \bar{x}^2}{4}-1}\right)\right]-\frac{5}{3}+4\gamma+2\log{\frac{\mu^2}{H^2}}  \right\} \ . \label{77}  }
\eea
\end{widetext}
We remark that, at leading order in the slow-roll expansion, this is an expression valid for small and large separations.
For scales larger than the Hubble horizon, $\Delta \bar{x}\gg 1$, we can further take the approximation, $4Re\left[ \log\left( \frac{\Delta \bar{x}}{2}+\sqrt{\frac{\Delta \bar{x}^2}{4}-1}\right)\right]\approx 2\log(\Delta \bar{x}^2)$.

\subsection{Angular power spectrum}

We now compute  the corresponding angular power spectrum from the modified two-point function for scalar perturbations and for low multipoles
\bea
C_{\ell}^{SW}=\frac{2\pi T_0^2}{25} \int_{-1}^{1}\langle {\cal {R}} (x) {\cal {R}} (x') \rangle_{ren}(y) P_{\ell}(y) dy \ . \label{61}
\eea
By construction this is a finite quantity, even without taking the large separation limit for the two-point function [as it was assumed in going from (\ref{16a}) to (\ref{17}) and (\ref{19b})].
To evaluate the logarithmic contributions of  (\ref{77}) to (\ref{61}) we take into account that $\int_{-1}^{1}dy \log(1-y) P_{\ell}(y)= - 2/\ell(\ell+1), \ell = 1, 2, \dots \ $.
The final result for the angular power spectrum with the modified two-point function is very well approximated by the following analytical expression:
\bea
C_{\ell}^{SW}& \approx &\frac{4\pi G}{\epsilon} \frac{8\pi T_0^2}{25} \frac{H^2(1-\epsilon)^2\bar r_L^{1-n}}{16\pi^2} \nonumber \\  &\times & \left\{   \frac{\Gamma(\ell+\frac{n-1}{2})}{\Gamma(\ell+2-\frac{n-1}{2})}-\frac{\bar r_L^{n-1}}{\ell(\ell+1)}\right\}      \ , 	\label{78}
\eea 
where we have used 
$\nu-\frac{3}{2}= \frac{1-n}{2}$, 
and $n$ represents the scalar index of inflation $n=1-4\epsilon-2\delta+O(\epsilon,\delta)^2$.
Also notice that expression (\ref{78}) is valid for $\ell\geq 1$, as for $\ell=0$ there would be present all the constant contributions  from the renormalized two-point function (\ref{77}), including the one depending on the renormalization scale. In fact, the renormalization scale may be fixed by imposing the natural condition $C^{SW}_{0}=0$. 

Notice that the first term in (\ref{78}) reproduces the standard result (\ref{19b}). The second one comes from the subtraction terms that we have added to the two-point correlator to continuously match the self-correlator at coincidence, but it shows scale invariance as well. Therefore, Eq. (\ref{78}) is consistent with observations
\cite{planck13}. 

However, the two terms in (\ref{78}) are competing, and the resulting amplitude for the coefficients $C_{\ell}^{SW}$  depends on the instant of time one evaluates $\bar r_L$. 
The first term is proportional to $H^2(t) \bar r_L^{(1-n)}(t)/\epsilon (t)$, and  it is time independent. However, the second term depends slightly on time.
The  value of $\bar r_L$ varies along the inflationary period, ranging  from $\bar r_L\approx 1$,  immediately after the instant of time $t_i$ at which the scale $r_L$ crosses the Hubble horizon  [$a(t_i) r_L \approx H^{-1}(t_i)$], to  $\bar r_L\approx e^{60}$, at the end of inflation (we have assumed that inflation lasts  for around $N=60$ $e$-foldings since the scale $r_L$ exited the horizon at $t_i$). In the former case, $\bar r_L\approx 1$, the amplitude is severely  reduced. In the latter situation,  $\bar r_L^{(n-1)} \sim 10^{-1}$, where we have assumed that  $n\approx 0.96$ \cite{planck13}, and  the amplitude is then  reduced at least  $10\%$. The adequate value of $\bar r_L$ to properly evaluate the resulting amplitude in (\ref{78}) is unclear. This question is closely related to the so-called ``quantum-to-classical transition" \cite{Liddle-Lyth00}, characterizing the period of time at which the primordial quantum perturbations behave as classical ones and define the initial conditions for the  postinflationary evolution, along with its associated power spectrum. In momentum space (mode-by-mode picture) this process is thought to happen a few Hubble times after horizon exit \cite{Liddle-Lyth00}, when the modes are frozen as classical perturbations. It seems natural to evaluate $\bar r_L$ during this period, where  quantum fluctuations are imprinted as classical perturbations. However, this quantum-to-classical mechanism is poorly understood, and it has not been rigorously established in the literature. Therefore one may regard $\bar r_L^{(n-1)}\equiv \alpha$ as a phenomenological parameter, varying in the range $1> \alpha > 0 $. Note that in the limiting case $\alpha \to 0$ one recovers the standard prediction, and this happens when the subtraction term is evaluated  after inflation. This parameter has influence  on  the relative strength between multipole amplitudes
\be 
\frac{\ell_2(\ell_2+1)C^{SW}_{\ell_2}}{\ell_1(\ell_1+1)C^{SW}_{\ell_1}}=\frac{ \ell_2(\ell_2+1)\frac{\Gamma(\ell_2+\frac{n-1}{2})}{\Gamma(\ell_2+2-\frac{n-1}{2})}-\alpha}{\ell_1(\ell_1+1)\frac{\Gamma(\ell_1+\frac{n-1}{2})}{\Gamma(\ell_1+2-\frac{n-1}{2})}-\alpha} \ . 
\ee
Observations may properly fit the value of this parameter.  It produces an observable effect for a significant range of values of $\alpha$. As remarked above, the details of how this ``quantum-to-classical transition'' takes place are not well established in the literature, and further work is needed to fully understand this process.
Within the present understanding of quantum gravity  it is difficult to determine theoretically the value of $\alpha$ and hence the relative impact of the subtraction term in the observed angular power spectrum.  However, as we showed above it can  potentially be tested with observations.

\section{Conclusions}
We have analyzed two-point correlators and self-correlators   of primordial perturbations in  quasi-de Sitter spacetime backgrounds.  For large separations two-point correlators exhibit nearly scale invariance in a very elegant way. We  have deformed the two-point correlators to smoothly match the self-correlators at coincidence.  To this end we have used renormalization methods in homogeneous backgrounds.  We have studied the physical consequences for the angular power spectrum at low multipoles. Scale invariance is maintained, but the amplitude of $C_{\ell}$  could change significantly. 
If one accepts a mismatch between the standard  two-point  correlators and the self-correlators and keeps only the large-scale behavior, the conventional predictions remain unaltered.  

We finally stress the importance of getting a better understanding of how to renormalize cosmological observables.  
The analysis carried out in the spacetime framework for  the tree-level power spectrum  may offer a way
to experimentally probe this issue.
\\

{\it Acknowledgements. } J. N-S. would like to thank I. Agullo, A. Fabbri,  
G. J. Olmo  and L. Parker for very useful discussions. This work is supported by the Spanish Grant No. FIS2011-29813-C02-02 and the Consolider Program CPANPHY-1205388.

\section*{Appendix}

In this appendix we give the basic steps to obtain the result (\ref{approx}). We consider (\ref{48}) first.
Since the first  prefactor is of order $O((\frac{3}{2}-\nu)^1)$,  we only need the corresponding hypergeometric function to be of order  $O((\frac{3}{2}-\nu)^0)$. One can see that
\small 
\bea
 Z^{-\frac{3}{2}-\nu}Re\left\{F\left(\frac{3}{2}+\nu,\frac{1}{2}+\nu,1+2\nu,\frac{1}{Z} \right)\right\}\Big|_{\nu=3/2} & = &\\
 6Re\left\{ \log(Z-1)\right\} -6\log(Z)+\frac{3}{Z}&-&\frac{3}{(1-Z)}\nonumber
\eea
\normalsize
On the other hand, the second prefactor of (\ref{48}) is of order $O((\frac{3}{2}-\nu)^0)$, so it is necessary to evaluate the second hypergeometric function at first order in the slow-roll series. To this end we will employ the following relation \cite{handbook}:
\bea
F\left(\frac{3}{2}-\nu,\frac{1}{2}-\nu,1-2\nu,\frac{1}{Z} \right) & = & \\
\left(1-\frac{ 1}{Z}\right)^{-3/4}P_{1/2}^{\nu}\left[ \frac{2Z-1}{2\sqrt{Z(Z-1)}}\right] &2^{-2\nu}&\Gamma(1-\nu)Z^{-\nu} \ , \nonumber
\eea
together with
\small
\bea
P_{1/2}^{\nu}(Z)=\left(\frac{Z+1}{Z -1} \right)^{\nu/2} \frac{F\left(-\frac{1}{2},\frac{3}{2},1-\nu,\frac{1-Z}{2}\right)}{\Gamma(1-\nu)} \ . 
\eea
\normalsize
At this point one can expand
\small
\bea
 F\left(-\frac{1}{2},\frac{3}{2},1-\nu,\frac{1-Z}{2}\right) & \approx & F\left(-\frac{1}{2},\frac{3}{2},-\frac{1}{2},\frac{1-Z}{2}\right)\\
  + \Big(\nu & - & \frac{3}{2}\Big) \frac{ dF\left(-\frac{1}{2},\frac{3}{2},1-\nu,\frac{1-Z}{2}\right) }{d\nu}\Big|_{\nu=3/2}  \nonumber
 \ , \eea
 \normalsize
 where the derivative can be performed using the representation series of the hypergeometric function.  Doing all the calculation properly one finally arrives at the following result:
\bea
Re\left\{F\left(\frac{3}{2}-\nu,\frac{1}{2}-\nu,1-2\nu,\frac{1}{Z} \right)\right\} & \approx & 1 \nonumber \\
+\Big(\frac{3}{2}-\nu\Big)  \left[  \frac{1}{4Z}+\frac{1}{4(1-Z)}-\frac{1}{2}Re\left\{ \log(Z-1)  \right\}   \right. & - & \frac{1}{2} \log(Z)  \nonumber \\
+ \left.  2 Re\left\{\log\left( \sqrt{Z}+\sqrt{Z-1}\right)\right\}   \right] &&  
\ . \eea
Taking all these results together for $Z\equiv \Delta \bar x^2/4$ in (\ref{48}) we can approximate the two-point function as in (\ref{approx}). We have also checked numerically that this expansion works well  irrespectively of the value of $Z$.


\begin{thebibliography}{99}
\bibitem{agullo-navarro} We note that for a free conformal-invariant field (like the electromagnetic one), the logarithmic  subtraction term is absent and hence also the renormalization  scale $\mu$. The two-point correlator is diluted by inflation and only the self-correlator survives as a quantity with non-negligible amplitude \cite{agullo-navarro13}.
\bibitem{note2} Alternatively, one could also consider short-distance modifications of the two-point function due to quantum gravity effects. See, for instance, Ref. \cite{padmanabhan}.
\end{thebibliography}
\end{document}